\documentclass[12pt,preprint]{aastex}

\shorttitle{Remnant Outflows from Massive Protostars}
\shortauthors{Klaassen \& Wilson}

\begin{document}

\title{Outflow and Infall in a Sample of Massive Star Forming Regions.}

\author{P. D. Klaassen \& C. D. Wilson}
\affil{Dept. of Physics and Astronomy, McMaster University, Hamilton, ON, Canada}
\email{klaassp@physics.mcmaster.ca}

\begin{abstract}
We present single pointing observations of SiO, HCO$^+$ and H$^{13}$CO$^+$ from the James Clerk Maxwell Telescope towards 23 massive star forming regions previously known to contain molecular outflows and ultracompact HII regions.  We detected SiO towards 14 sources and suggest that the non-detections in the other nine sources could be due to those outflows being older and without ongoing shocks to replenish the SiO.  We serendipitously detected SO$_2$ towards 17 sources in the same tuning as HCO$^+$. We detected HCO$^+$ towards all sources, and suggest that it is tracing infall in nine cases.  For seven infall candidates, we estimate mass infall rates between 1$\times10^{-2}$ and 2$\times10^{-5}$ M$_{\odot}$ yr$^{-1}$.  Seven sources show both SiO detections (young outflows) and HCO$^+$ infall signatures. We also find that the abundance of H$^{13}$CO$^+$ tends to increase along with the abundance of SiO in sources for which we could determine abundances.  We discuss these results with respect to current theories of massive star formation via accretion. From this survey, we suggest that perhaps both models of ionized accretion and halted accretion may be important in describing the evolution of a massive protostar (or protostars) beyond the formation of an HII region.

\end{abstract}


\section{Introduction}
\label{sec:intro}

The  dynamics in massive star forming regions are, in general, much more complex than in regions which form only low mass stars.  For instance, in the early stages within a low mass star forming region, the dynamics can be understood in terms of a few broad categories: large scale infall, which causes a disk to form, accretion through the disk, and outflow to release angular momentum (see for example, Di Francesco et al. 2001, Andr\'e et al. 1993, Muzerolle et al. 2003). In intermediate and high mass star forming regions, turbulence, stellar winds, multiple sites of star formation, and, for regions with massive star formation, the presence of HII regions, all contribute to the dynamics in these regions as well (i.e. Beuther et al. 2006, Shepherd \& Churchwell 1996, McKee \& Tan 2003, Krumholz et al. 2005). The more complicated source dynamics make the processes involved in the formation of the most massive stars much more difficult to understand than those involved in the formation of lower mass stars. Adding to the complexity, massive stars do not form as often as their lower mass counterparts and so we must look to larger distances before finding examples of high mass star formation.  For instance, the average distance to the 63 sources in Shirley et al. (2003, hereafter S03) is 5.3 kpc.

  If we assume  massive stars form through accretion, that this accretion occurs in the inner regions of disks (i.e. Pudritz \& Norman 1986, or Shu et al. 1994), and that these disks have radii of a few thousand AU (Chini et al. 2004, Beltran et al. 2004, Cesaroni et al. 2005), we do not yet quite have the resolving power to detect accretion directly at the distances to massive star forming regions (1000 AU at 5.3 kpc is $\sim0.2''$). Here, we define accretion as the infall motions from the disk onto the forming star, in contrast to the larger scale motions of envelope material falling onto the disk. However, while we cannot observe accretion directly, its presence can be inferred from the presence of accretion tracers such as larger scale infall and outflow. Infall can act to replenish disk material as mass accretes onto a protostar (Nakamura 2000), while molecular outflows serve as a release mechanism for the angular momentum which builds up during the accretion process (e.g Arce et al. 2006). These large scale motions are seen in star forming regions of all mass scales (see for instance Beuther \& Shepherd 2005).

In massive star forming regions, the accretion rates are orders of magnitude greater than in low mass star forming regions (i.e. Beuther et al. 2002), while the accreted masses are only approximately one order of magnitude greater. These accretion rates and masses result in accretion timescales that are much shorter than in low mass star forming regions, which allows the Kelvin-Helmholtz timescale to become important in the evolution of the protostar (e.g., there is no pre-main sequence stage for massive star formation).  The outward radiation and thermal pressure from the forming star becomes strong enough that it can ionize the surrounding medium and a small, highly ionized HII region (either hypercompact (HCHII) or ultracompact (UCHII) region, Keto 2003) can form.  It is still unclear whether the outward pressure needed to create the HII region is strong enough to halt accretion, or whether accretion can continue in some form (either through a molecular or ionized disk, or through an ionized accretion flow) after the formation of an HII region.  Some models suggest that accretion must halt  before the onset of a visible UCHII region (i.e. Garay \& Lizano 1999, Yorke 2002), while other models suggest that an ionized accretion flow can continue through an HII region (i.e. Keto 2003, 2006). 

There is now also observational evidence which suggests that, once the protostar becomes hot enough to ionize its surroundings, both modes of massive star formation (halted and ionized accretion) are possible. G10.6-0.4 has been shown to have an ionized accretion flow by Sollins et al. (2005) and Keto \& Wood (2006), while accretion in G5.89-0.39 seems to have halted at the onset of the UCHII region (Klaassen et al. 2006). Although sample statistics at this point are still quite small, these two examples pose interesting questions. We do not yet have enough data to determine whether the apparently conflicting models of halted and ionized accretion can both be correct.  However, we can begin with a uniform survey of infall and outflow tracers in massive star forming regions in order to constrain massive star formation scenarios.

In this paper, we present a survey of 23 massive star forming regions. Because we are interested in the relationship between accretion and outflow after the formation of an HII region, our source selection criteria include (1) the presence of an UCHII region, which indicates that there is a massive protostar forming, and (2) previous evidence of outflows, which suggests ongoing accretion in most formation scenarios.  Sources were selected based on inclusion in the Wood \& Churchwell (1989) and Kurtz et al. (1994) catalogs of UCHII regions as well as having molecular outflow signatures in the Plume et al. (1992) survey of massive star forming regions.  Additional sources were taken from Hunter (1997) which were shown to have both UCHII regions and molecular outflows.

We describe the observations collected for this survey in Section \ref{sec:observations}, we discuss the results of these observations in Section \ref{sec:results}, and present our conclusions in Section \ref{sec:conclusions}.

\section{Observations}
\label{sec:observations}

Observations of SiO (J=8-7), HCO$^+$, and H$^{13}$CO$^+$ (J=4-3) were taken at the James Clerk Maxwell Telescope (JCMT)\footnote{The James Clerk Maxwell Telescope is operated by The Joint Astronomy Centre on behalf of the Particle Physics and Astronomy Research Council of the United Kingdom, the Netherlands Organisation for Scientific Research, and the National Research Council of Canada.} in 2005 (as parts of projects M05AC11 and M05BC04).  SiO (347.330 GHz) and H$^{13}$CO$^+$ (346.999 GHz) were observed simultaneously in the same sideband by tuning the receiver to 347.165 GHz.  Thirteen or twenty minute observations, depending on the source elevation, were taken towards each source with a velocity resolution of 1.08 km s$^{-1}$, which resulted in rms noise levels of T$_{\rm MB}<$0.07 K. Separately, we observed  HCO$^+$ (356.370 GHz) with a velocity resolution of 0.53 km s$^{-1}$ to an rms noise limit of T$_{\rm MB}<$0.13 K in twenty minute integrations.  Both sets of observations were taken in position switching mode with dual mixers and a sideband rejection filter in place. Table \ref{tab:sources} shows the positions, rms noise limits of both tunings, the local standard of rest velocity, and distances to all sources in this survey. The half power beam width for these observations is 15$''$, and the main beam efficiency is $\eta_{\rm mb}$=0.62. Data were obtained using the DAS autocorrelator system and reduced using the SPECX software package.

Linear baselines were removed from all spectra except for those towards G10.47.  For this source, there were so many different chemical species in the observed spectrum that we were unable to fit a linear baseline over the entire $\sim$ 700 km s$^{-1}$ bandwidth of the 347 GHz observations or the $\sim$ 450 km s$^{-1}$ bandwidth of the 356 GHz observations.  In this case, no baseline was removed.
 
We also present $9\times9$ maps of one source (G45.07) in the same emission lines.  These raster maps are sampled every 5'' and have rms noise limits of 0.10 and 0.14 K (T$_{\rm MB}$) for the 347 and 356 GHz tunings, respectively.  Note that these values are different than the ones reported in Table \ref{tab:sources} for the single pointing observations. The DAS autocorrelator was configured with the same tunings as were described above for the single pointing observations, and the maps were centered at the same position.

\section{Results}
\label{sec:results}

This survey of single pointing observations towards 23 massive star forming regions is meant as an initial, uniform survey from which to base future observations.  With these observations, we can only comment on the molecular gas component within our beams; we cannot discuss the larger scale molecular dynamics, or the ionized gas components of these regions.  For our sources, the selection criteria of having an HII region confirms that these regions are forming massive stars. 

\begin{figure}
\includegraphics[scale=0.7,angle=-90]{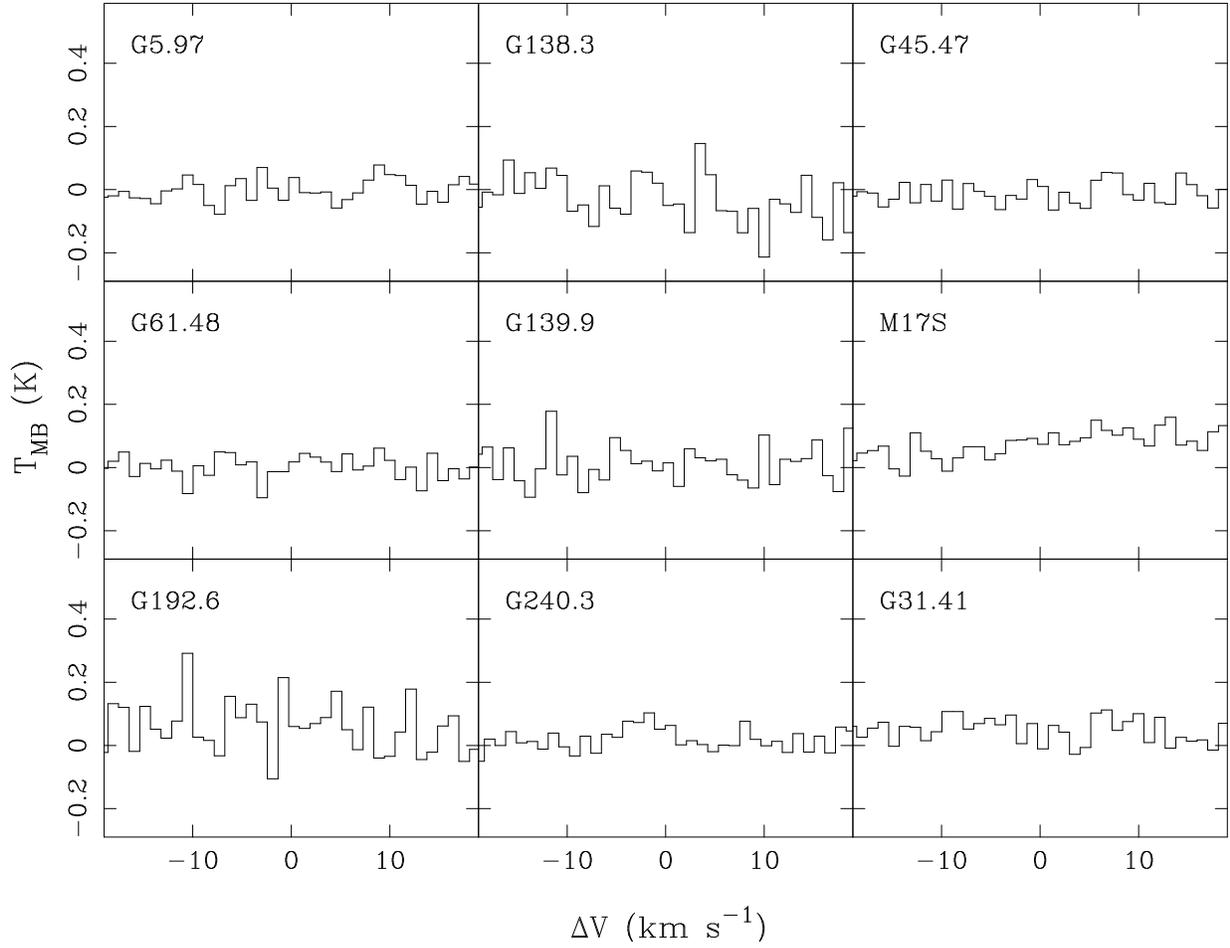}
\caption{Nine sources in which SiO was not detected (to 4 $\sigma$ limits). The name of each source is given in the top left hand corner of each panel. For each source, the source rest velocity is plotted as $\Delta v$ = 0 km s$^{-1}$.}
\label{fig:Sno_sio}
\end{figure}

\begin{figure}
\includegraphics[scale=0.7,angle=-90]{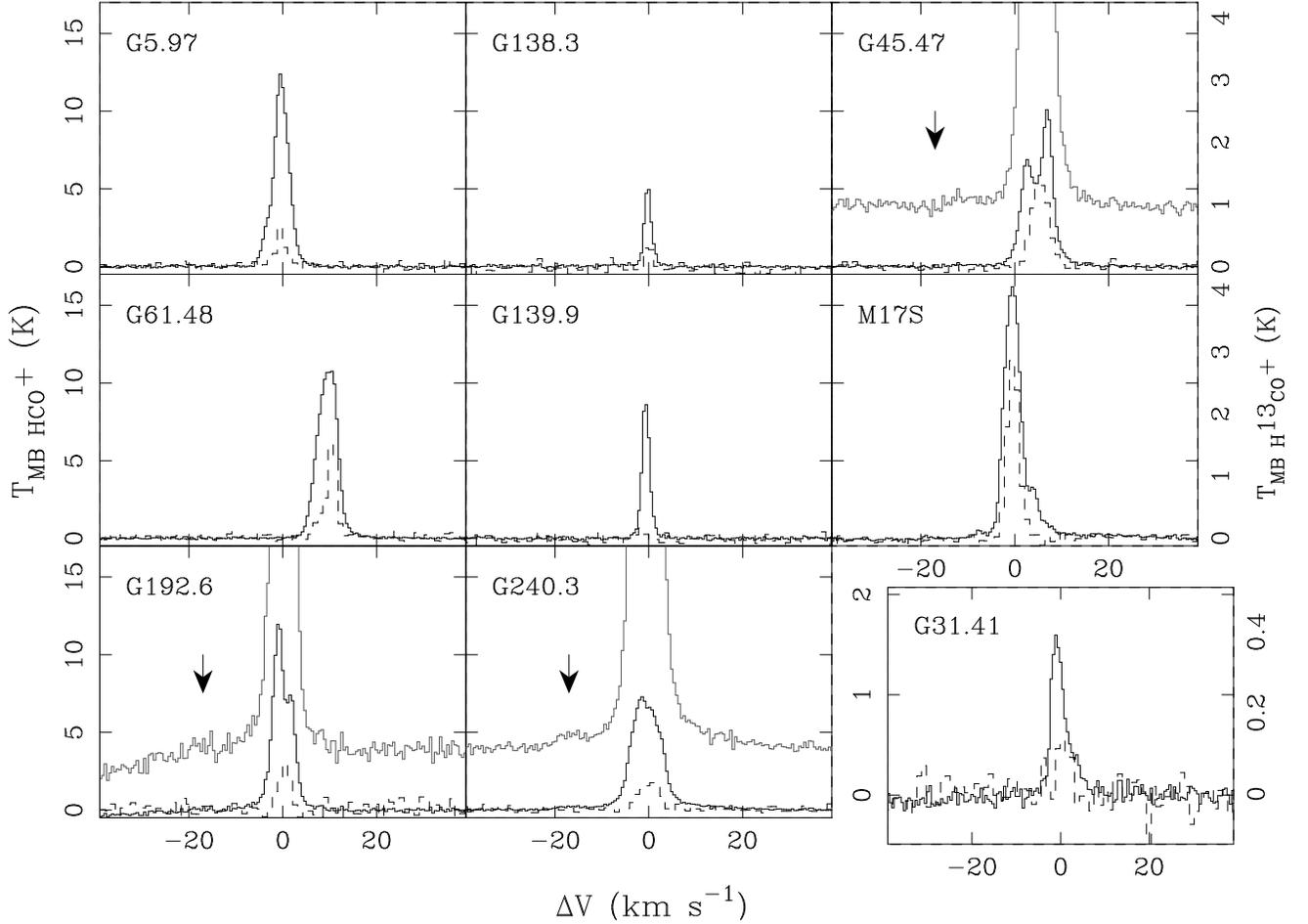}
\caption{HCO$^+$ and H$^{13}$CO$^+$ sources with no SiO detections (the same sources as in Figure \ref{fig:Sno_sio}).  Solid lines show HCO$^+$ emission, while dashed lines show H$^{13}$CO$^+$ emission scaled up by a factor of four. The temperature scale on the left hand side of the panels is the scale used for the HCO$^+$ spectra, and the temperature scale on the right hand side is that used for the H$^{13}$CO$^+$ spectra. For sources with SO$_2$ detections, we have re-plotted the HCO$^+$ spectra on a larger intensity scale ( from -1 K to 3 K, in gray) to show the low lying SO$_2$ emission. The SO$_2$ emission, when present, is centered at  -17 km s$^{-1}$, and is indicated by an arrow. G31.41 is slightly offset to emphasize that the temperature scale has been magnified to show the emission feature.}
\label{fig:Hno_sio}
\end{figure}

\begin{figure}
\includegraphics[scale=0.7,angle=-90]{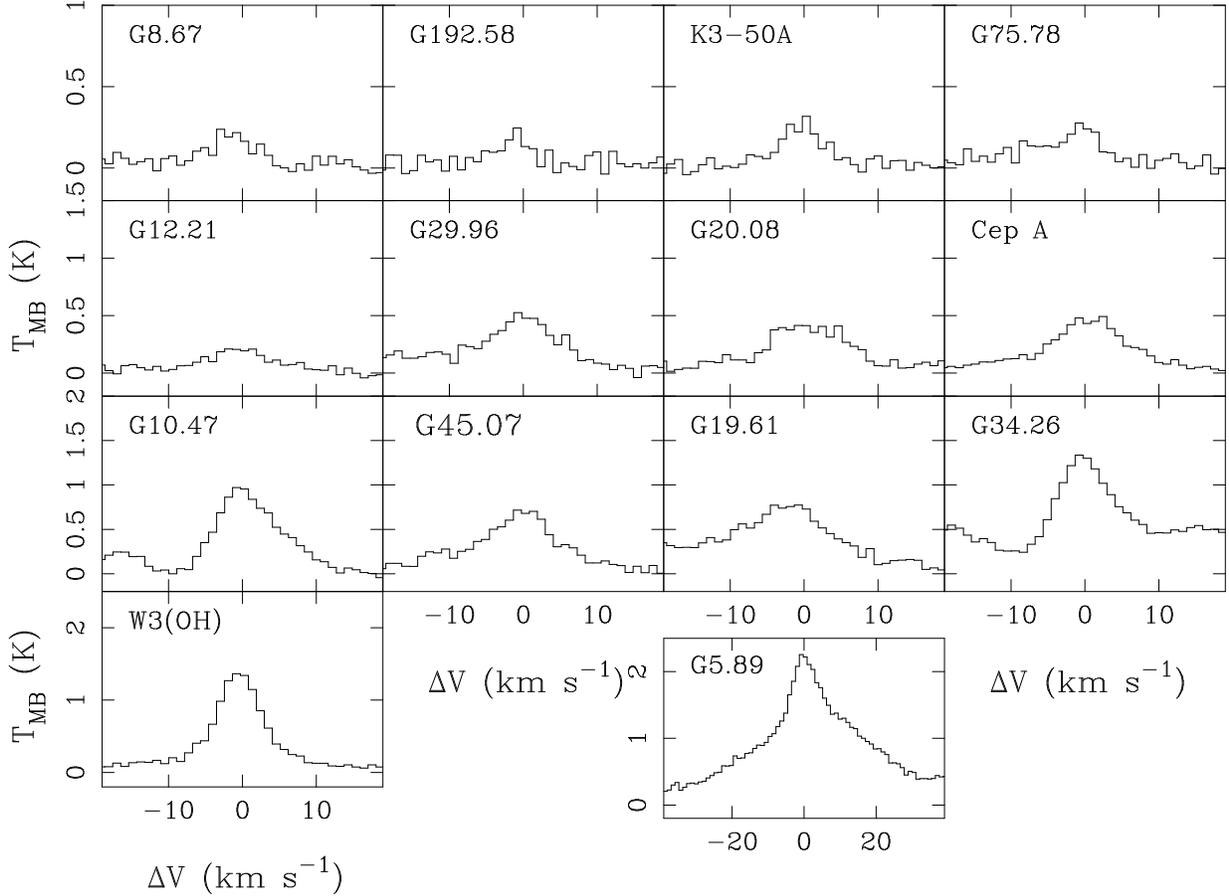}
\caption{SiO spectra for 14 source in which SiO was detected (above 4 $\sigma$ limits). For this figure only, the range of temperatures plotted increases by row. The plotted temperature ranges for each row increase by 0.5 K per row.  The temperature scale in the top row extends from 0.2 to 1 K (in the $T_{\rm MB}$ scale), while the temperature scale in the bottom row extends from 0.2 to 2.5 K. Note that the velocity scale for G5.89 is much wider (a 78 km s$^{-1}$ window as opposed to a 38 km s$^{-1}$ window for the other sources) in order to show the full width of the emission line, but that the temperature scale is the same as that for W3(OH).  }
\label{fig:Sdet_sio}
\end{figure}

\begin{figure}
\includegraphics[scale=0.7,angle=-90]{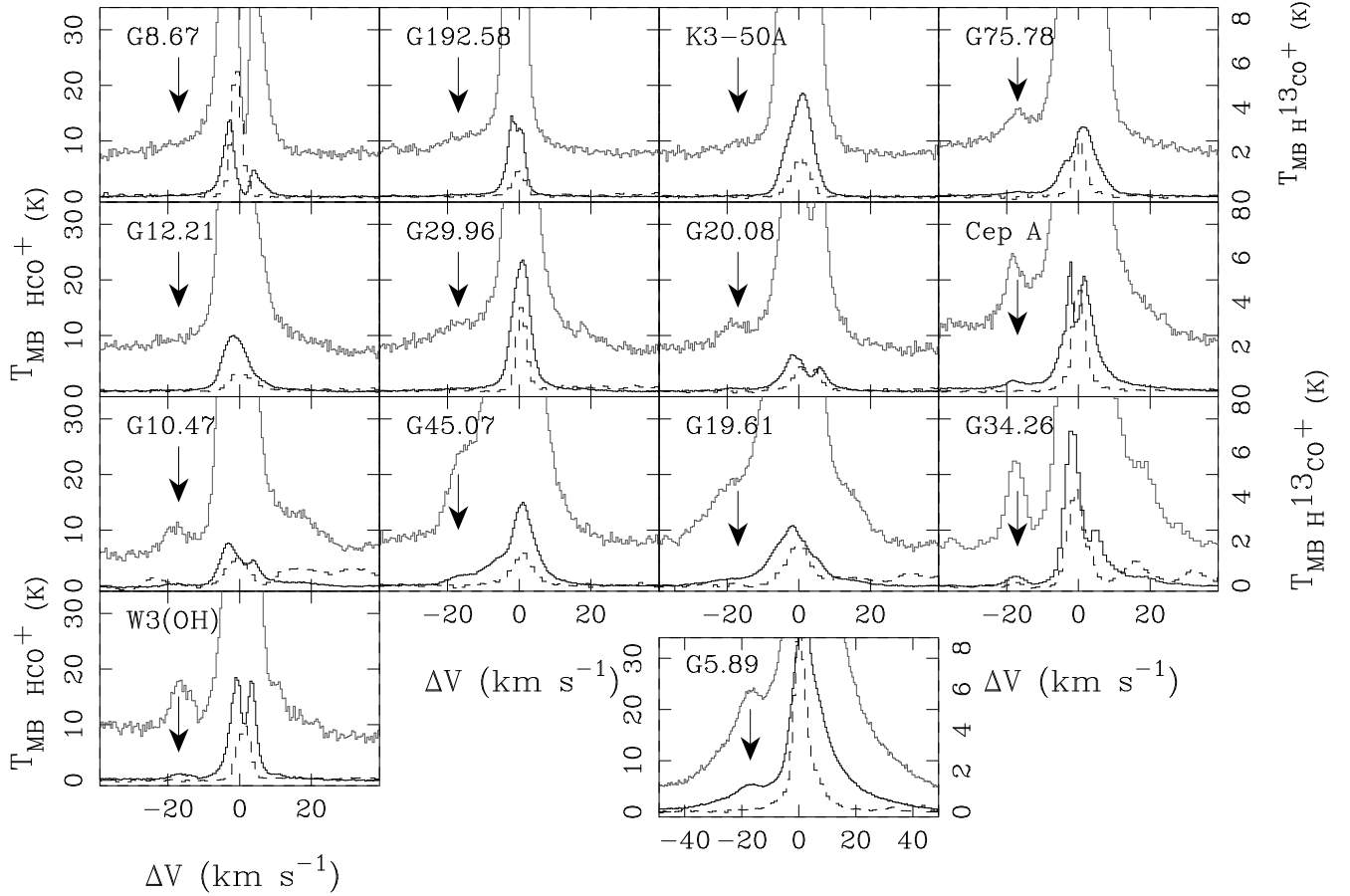}
\caption{HCO$^+$ and H$^{13}$CO$^+$ emission from sources with SiO detections (the same sources as in Figure \ref{fig:Sdet_sio}).  The solid, dashed and gray lines are the same as shown in Figure \ref{fig:Hno_sio}, as is the placement of the temperature scales. As in Figure \ref{fig:Sdet_sio}, G5.89 has been plotted separately from the rest of the sources to stress that the velocity scale is larger for this source. }
\label{fig:Hdet_sio}
\end{figure}

Figures \ref{fig:Sno_sio} through \ref{fig:Hdet_sio} show the single pointing SiO and HCO$^+$/H$^{13}$CO$^+$ spectra towards all sources; the spectra are ordered according to SiO integrated intensity. For each panel in these figures, line brightnesses have been corrected for the JCMT main beam efficiency and centered on the $V_{\rm LSR}$ of the source (Table \ref{tab:sources}).  Figures \ref{fig:Sno_sio} and \ref{fig:Hno_sio} show SiO and HCO$^+$/H$^{13}$CO$^+$ spectra, respectively, towards the sources with no SiO detections. Figures \ref{fig:Sdet_sio} and \ref{fig:Hdet_sio} show the SiO and HCO$^+$/H$^{13}$CO$^+$ spectra, respectively, towards sources with SiO detections.  Peak line strengths and integrated intensities are given for all three lines in Table \ref{tab:fits}.

 SiO  was only detected in 14 out of our 23 sources, where we define a detection as a minimum of 4 $\sigma$ in integrated intensity.  The rms noise limits in integrated intensity were calculated using $\Delta I = T_{\rm rms}\Delta v\sqrt{N_{\rm chan}}$ where $T_{\rm rms}$ is the rms noise level in K, $\Delta v$ is the velocity resolution of the observations, and $N_{\rm  chan}$ is the number of channels over which the integrated intensity is calculated.  HCO$^+$ was detected in all sources  and H$^{13}$CO$^+$ was detected in all but two sources. Along with HCO$^+$, we serendipitously observed SO$_2$ (J=10$_{4,6}$-10$_{3,7}$ at 356.755 GHz) in 17 of our sources. The HCO$^+$ and SO$_2$ lines are only separated by 17 km s$^{-1}$ and thus the lines were blended in eight sources. For the sources with SO$_2$ detections, we have also plotted (in gray) the lower intensity HCO$^+$ observations in order to highlight the SO$_2$ emission (Figures \ref{fig:Hno_sio} and \ref{fig:Hdet_sio}).

 Double peaked HCO$^+$ line profiles were observed towards 10  sources, with nine of them having stronger blue peaks than red.  This blue line asymmetry in an optically thick tracer such as HCO$^+$ is often suggestive of infall (i.e Myers et al. 1996). We discuss the possibility of our observations tracing large scale infall further in Section \ref{sec:hcop}.

The distances to our sources, as taken from the literature, are shown in Table \ref{tab:sources}.  The average distance is 5.7 $\pm$ 3.8 kpc, where the error quoted reflects the 1 $\sigma$ dispersion in the distances. Since our observations were taken with a 15$''$ beam, this resolution corresponds an average linear size of 0.4 pc for our observations.    

SiO is a well known outflow tracer, since in the general interstellar medium, Si is frozen out onto dust grains.  When the gas in a region is shocked (i.e. the gas through which a protostellar outflow is passing) the dust grains can sublimate and Si is released into the gas phase.  After the passage of a shock, the SiO abundance ([SiO]/[H$_2$]) can jump to almost 10$^{-6}$, whereas the dark cloud abundance of SiO is often closer to 10$^{-12}$ (see for example Schilke et al. 1997, Caselli et al. 1997, or van Dishoeck \& Blake, 1998).

While SiO is easily identified as an outflow tracer, the emitting region for HCO$^+$ is much less certain.  Many authors suggest that HCO$^+$ can be used to trace the envelope material surrounding a protostellar region (i.e. Hogerheijde et al. 1997, Rawlings et al. 2004), while others suggest that it traces disk material (i.e. Dutrey et al. 1997).  One thing that is apparent, however, is that it becomes optically thick very quickly and  readily self absorbs.

We detected SO$_2$ in 17 of our sources, suggesting that our beam contains at least some molecular gas at temperatures greater than 100 K (see for instance Doty et al. 2002, Charnley 1997). However, Fontani et al. (2002) have determined the average temperature in twelve massive star forming regions to be 44 K, using observations with beam sizes comparable to those presented here. For seven of our sources, which were observed in the Fontani et al (2002) sample, the average temperature  is also 44 K. Thus, in the following analysis, we adopt an ambient temperature of 44 K for all sources.

Table \ref{tab:results} shows the column densities for each region derived for both SiO and H$^{13}$CO$^+$. The column density was calculated assuming that each tracer is optically thin, in local thermodynamic equilibrium, and at an ambient temperature of 44 K.  For optically thin lines, the column density of the observed transition scales directly with the integrated intensity of the line (see for instance, Tielens 2005):

\begin{equation}
N_u = \frac{8k\pi\nu^2}{hc^3}\frac{1}{A_{u\ell}}\int T_{\rm MB} dv
\end{equation}

\noindent where $N_u$ is the column density in the upper state of the transition, $A_{u\ell}$ is the Einstein A coefficient, $\nu$ is the frequency of the J=$u$-$\ell$ transition, and $\int T_{\rm MB}dv$ is the integrated intensity of the line.  The column density of this one state can then be related to the total column density of that molecule through the partition function. It is the total column density for the molecule (not the observed state) that is presented in Table \ref{tab:results}.

\subsection{Source Properties derived from SiO observations}
\label{sec:SiO}

For each source we determined the column density, or upper limit to the column density, in SiO (Table \ref{tab:results}) using the methods described above.  These column densities can be compared to the column densities of other molecules  (i.e. CS) for the same regions in order to determine the fractional abundance of SiO, if the abundance of the other molecule is known.  We were able to obtain the CS or C$^{34}$S column densities for fifteen of our sources from the literature. Column densities for fourteen sources were taken from Plume et al. (1997, hereafter P97), with the column density for one additional source taken from Wang et al. (1993). For those sources with C$^{34}$S column densities instead of CS column densities, we assumed an abundance ratio of [CS]/[C$^{34}$S] = 22 (Wilson \& Rood, 1994) to determine a CS column density.  The abundance of CS, relative to H$_2$, was calculated by S03 for 13 of these sources, and we assume a CS abundance of $1.2\times10^{-9}$ for the other two source for which the CS column density is known, since this was the average CS abundance as calculated by S03. We then compare the column density and abundance of CS to our observed SiO column density, or column density upper limit, to determine the abundance of SiO relative to H$_2$ in our sources (Table \ref{tab:results}).

Despite our source selection criteria requiring previous evidence of outflows, we detected SiO towards only 14 of our 23 sources. This raises a number of questions, such as: is the observed SiO in fact tracing outflow if we do not detect it in all sources? Why do we not detect SiO in all sources? Is the signal being beam diluted at large distances? Has the Si evolved into other species?  Below we first address whether the detected SiO can be used as an outflow tracer, and then discuss reasons for our non-detections of SiO in nine of our sources.

Si is liberated in shocks, and if these shocks are not due to the outflow, they must be due to the photo dissociation region (PDR) surrounding the UCHII region.  Evidence for diffuse (not collimated) SiO can be seen in W75N (Shepherd, Kurtz \& Testi, 2004) suggesting that the SiO may be due to the PDR and not an outflow. Models and observations of SiO in the PDRs around high mass star forming regions suggest moderate SiO enhancement, and that the SiO abundance is independent of the ambient radiation field (i.e. Schilke et al. 2001).  Schilke et al. (2001) find SiO column densities of $\sim10^{12}$ cm$^{-2}$ in their observed PDRs. This is, admittedly, below our detection threshold; however, we detect average SiO column densities of $\sim 10^{14}$ cm$^{-2}$. This suggests a possibly higher SiO abundance than found in PDRs.

\begin{figure}
\includegraphics[scale=0.7,angle=-90]{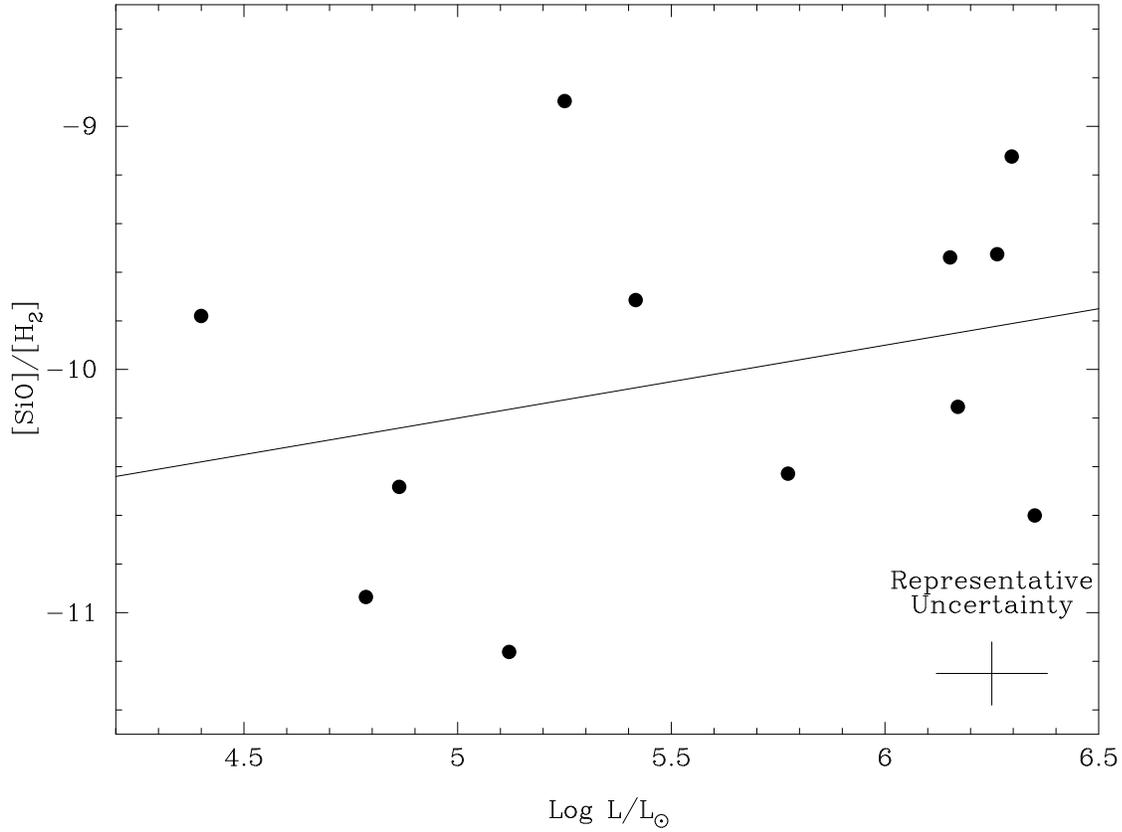}
\caption{SiO abundance plotted as a function of source luminosity.  The line of best fit shown is given by (Log([SiO]/[H$_2$])=0.30$\pm$0.06Log($L/L_{\odot}$)-11.7$\pm$0.3), and was calculated only for sources detected in SiO.  This shows that SiO abundance increases with source luminosity. This is contrary to what is expected for SiO produced in PDRs, suggesting the observed SiO is produced in outflows.}
\label{fig:iras}
\end{figure}

The enhanced SiO column density alone is not enough to discount the origin of the SiO in our sources as the PDR and so we can consider how our SiO abundance varies with the ambient radiation field.   For the gas near an HII region, we can approximate the strength of the ambient radiation field using the Far Infrared (FIR) luminosity of the region. For twelve of our sources with SiO detections and abundance calculations, we obtained the FIR luminosity from either Wood \& Churchwell (1989), Kurtz et al. (1994) or Evans et al. (1981). We then scaled their values for the different source distances used in this study (see Table \ref{tab:results}). Comparing the SiO abundance to the source luminosity (see Figure \ref{fig:iras}), we find that the SiO abundance increases with source luminosity. There is a 10\% chance that this relationship could arise from uncorrelated data.  Thus, it is possible that our result is contrary to the findings of Schilke et al (2001), and we suggest that the SiO we observe does come primarily from outflow shocks.

We can also compare our detection rate of SiO to that of the SiO survey towards maser sources of Harju et al (1998) who observed SiO (J=2-1) and SiO (J=3-2). For comparison to our results, we only consider the sources in Harju et al. which are listed as UCHII regions.  Our detection rate is 61\%, compared to their rate of 29\%. While our detection threshold is slightly lower (our observations have rms noise levels generally below 0.05 K at 347 GHz, while their rms noise levels are generally below 0.08 K), we suggest that the different detection rates are due to differences in the source selection criteria.  Although both samples contain UCHII regions, our sample contains sources with previous observations of outflows, while Harju et al. have selected sources based on previous observations of masers.  For the 12 sources which overlap between the two studies, we detect SiO towards 9 sources, while they detect SiO towards 10. They detected SiO (J=2-1) in G31.41 while we did not detect it in SiO (J=8-7).

Based on this comparison to SiO observations of UCHII regions {\it not} selected by outflows, which have a lower SiO detection rate, and that our SiO abundance increases with source luminosity, we suggest that the SiO, in the 14 sources in which it is detected, is being generated in the outflow. Previous, high resolution observations of SiO also suggest that SiO can be enhanced in the outflows from high mass stars (i.e. Beuther et al. 2004, Beuther, Schilke \& Gueth 2004) just as it is in the outflows from low mass stars.

 The  nine non-detections in our sample could be caused by beam dilution if these sources are on average further away. However, if we compare the distances for sources with and without SiO detections, we find average distances of 6.3 $\pm$ 4.4 and 4.4 $\pm$ 2.5 kpc, respectively. Thus, the non-detections cannot be attributed to larger average source distances and so beam dilution can play only a minimal role in the non-detections of SiO.

If these SiO non-detections are not due to distance effects, there must be some local phenomenon which can explain why SiO is not being detected in regions known to contain protostellar outflows.  It is possible that the Si is evolving into different species and the SiO abundance is dropping back down to dark cloud values.   Pineau des For\^ets et al. (1997) suggest that a few$\times10^4$ yr after the Si is liberated from dust grains and forms SiO, it can either freeze out back onto dust grains or oxidize and form SiO$_2$. Thus, the lack of SiO may be due to silicon moving into other species if it was liberated more than 10$^4$ yr ago.  This interpretation implies that we did not detect SiO in some of our sources because the outflow generating mechanism shut off more than 10$^4$ years ago, and the outflow observed in HCO$^+$ (or in other molecules by other authors) is a remnant of previous accretion.

\begin{figure}
\includegraphics[scale=0.7,angle=-90]{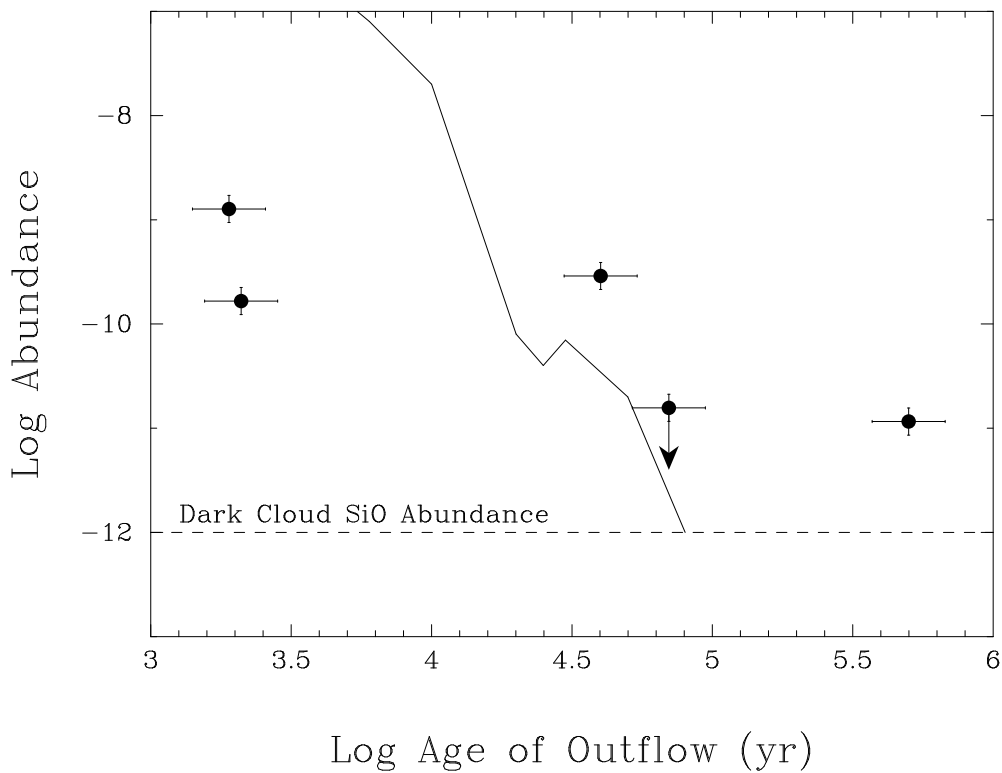}
\caption{SiO abundance plotted as a function of  outflow age.  The outflow ages were taken from Wu et al. (2004).  The solid line shows the model predictions of Pineau des For\'ets et al. (1997) for the SiO abundance as a function of age (assuming Si returns to the dust grains), and the dashed line represents the canonical dark cloud abundance of SiO. The source shown with a downwards arrow represents G61.48, a source in which we did not detect SiO, and the value given is an upper limit to the SiO abundance. The error bars represent 30\% calibration uncertainty between our observations and those of P97 and S03}
\label{fig:age_abund}
\end{figure}

 The kinematic ages of nine of our sources are listed in the Wu et al (2004) catalog of high velocity outflows.  Of these nine sources, five were also included in the P97 and S03 studies. This results in five sources for which we have both the kinematic age of the outflow, and the abundance of SiO.  The relationship between outflow age and SiO abundance is shown in Figure \ref{fig:age_abund} along with the model predictions of Pineau des For\^ets et al. (1997). With only five points, it is difficult to draw conclusions about the relationship between outflow age and SiO abundance, especially given the uncertain beam filling factor.  At higher resolution, these points would likely move upwards to higher abundances.  A larger beam filling factor would move the points for the two young outflows (G5.89 and Cep A) towards the model predictions.  As for the other three outflows, this would move them further from the model predictions.  We suggest that the outflow generating mechanism is continuing to shock these regions, replenishing the SiO. The oldest of these sources (G192.58) shows an infall signature in HCO$^+$, which also suggests that the outflow is still being powered.

\subsection{Source properties derived from HCO$^+$ observations}
\label{sec:hcop}

Given the large average distance to our sources (5.7 kpc), our 15$''$ beam subtends an average linear distance of 0.4 pc.  Thus, it is quite likely that the HCO$^+$ emitting region does not entirely fill our single JCMT beam.  In general we can determine the beam filling factor, $f$, for each source using 

\begin{equation}
T_L = f_{\rm source}T_s(1-e^{-\tau})
\label{eqn:filling_factor}
\end{equation}

\noindent where $T_L$ is the line brightness temperature measured at the telescope (corrected for telescope efficiencies) and $T_s$ is an approximation to the ambient temperature (44 K) which is valid at densities greater than $\sim10^3$ cm$^{-3}$ (Rholfs \& Wilson, 1994). In a number of our sources, the HCO$^+$ is asymmetric, and thus cannot be consistently used to determine a beam filling factor.  Instead, we can use the optically thin H$^{13}$CO$^+$ line, and we can simplify the above equation to $f=T_L/(T_s\tau)$. These values are shown in Table \ref{tab:results}.

HCO$^+$  becomes optically thick quite quickly due to its relatively high abundance with respect to H$_2$ ([HCO$^+$]/[H$_2$] $\sim 10^{-8}$), and as such, can be used to roughly trace outflow and to trace infall (i.e. Myers et al. 1996) if the line profile shows a double peak.  We determined the optical depth of HCO$^+$ towards each source using Equation 1 of Choi et al. (1993), and found that in all but one case (G139.9), it is optically thick. In all cases, the optical depth of HCO$^+$ is less than 77 (the abundance ratio between HCO$^+$ and H$^{13}$CO$^+$, Wilson \& Rood, 1994), resulting in optically thin H$^{13}$CO$^+$ towards all sources. Because HCO$^+$ is optically thick in its line center, the line wings can be used to detect outflows, and so, if there is an outflow, it should be detectable in HCO$^+$ even if it goes undetected in SiO.  Gaussian profiles were fit to our HCO$^+$ spectra (either single or double Gaussians, depending on the observed line shape), and the fits were subsequently subtracted from the spectra to leave only the residual outflowing gas. When using two Gaussians to fit the self absorbed spectra, we employed a method similar to the single Gaussian fitting of Purcell et al. (2006) because we used the sides of the detected lines to fit our profiles (see their Figure 3).  Comparing the two Gaussian fits to single Gaussian fits showed no significant differences in distinguishing line wing intensity. Because of the possibility of contamination from SO$_2$ emission in the blue shifted outflow wing (at -17 km s$^{-1}$), the peak brightness of the residual emission was determined using only the red shifted wing emission.  In all cases (except for G10.47 for which we could not find a linear baseline), we found a minimum of a 5$\sigma$ peak brightness temperature in the residual line wing emission, with 19 of our sources having a minimum of a 10$\sigma$ peak.  This result suggests that we can detect outflow motions in all sources using our detections of HCO$^+$, despite not detecting SiO towards every source.

  Our observations show that for ten of our sources, the spectral line profile of the HCO$^+$ emission has a double peak.  This profile could either be due to self absorption of the optically thick HCO$^+$ line or from multiple velocity components within our 15$''$ beam. To break this degeneracy, we observed  the optically thin H$^{13}$CO$^+$. If the H$^{13}$CO$^+$ line has a single peak at the same velocity as the HCO$^+$ absorption feature, then it is likely that the HCO$^+$ line is self absorbed. If, however, the H$^{13}$CO$^+$ also has two peaks, and they are at approximately the same velocities as the two HCO$^+$ peaks, it would suggest that there are multiple components within the beam. Of our ten sources with double HCO$^+$ peaks, only one shows a double peak profile in H$^{13}$CO$^+$ (G20.08). This results in nine sources with double peaked optically thick HCO$^+$.

In addition to the nine optically thick sources, similar line asymmetries appear in a number of other sources.  However, in these sources, there is no clear emission gap producing a double peak profile, only an emission shoulder (i.e. De Vries \& Myers 2005).  If we take G75.78 as an example, the HCO$^+$ line peak is red shifted from the rest velocity of the source, with a blue shifted emission shoulder.

There are a number of different kinds of source dynamics that can lead to the double peaked line profiles seen in our spectra, such as infall, outflow and even rotation.  However, infall is the only one of these processes which would produce line asymmetries which are consistently blue (i.e. the blue peak is higher than the red peak or shoulder).  If these profiles were due to outflow or rotation, there would be no statistical reason to have more sources with higher blue peaks than red peaks. Many previous studies have investigated the statistical significance of using this type of optically thick blue line asymmetry to trace infall as opposed to other dynamical motions (i.e. Mardones et al. 1997 and Gregersen et al. 1997 for low mass star forming regions, and Fuller et al. 2005 for high mass star forming regions). 

Of the 10 sources in our survey which have double peaked HCO$^+$ profiles, we suggest eight may be indicative of infall.  The other two sources are G20.08 and G45.47. G20.08 has already been shown to have multiple components in the beam from the double peaked H$^{13}$CO$^+$ profile, and G45.47 has a brighter red peak than blue. There are two additional sources (G19.61, and G240.3) in which HCO$^+$ has a strong red shifted shoulder, which we suggest may also be tracing infall. This analysis gives a total of ten infall candidates in our sample of 23 sources.

A recent survey of HCO$^+$ (J=1-0) towards sources with methanol masers shows an even distribution of sources with blue and red line asymmetries, and a higher percentage of self absorbed lines than in our study (Purcell et al. 2006).  Of the six sources which overlap between our survey and that of Purcell et al, all six are self absorbed in HCO$^+$ (J=1-0). Five of them have blue line asymmetries consistent with infall, while only one (G31.41) has its red peak brighter than its blue peak.  We only find self-absorption in HCO$^+$ (J=4-3) for three of these six sources.  In two of the sources for which we do not see a clear self-absorption feature, we do see evidence for a red shifted shoulder which may be showing unresolved infall.

The sources in Purcell et al. (2006) have an even distribution of red and blue line asymmetries, while we have a clear bias towards detecting blue line asymmetries.  This comparison could suggest that the higher energy J=4-3 transition of HCO$^+$ is a better tracer of infalling gas because it does not self absorb as readily as the J=1-0 transition.

For each of the 8 sources with blue, double peaked HCO$^+$ profiles, we can determine an infall velocity ($v_{\rm in}$) using the two layer radiative transfer model of Myers et al. (1996).  Using their equation 9, we find infall velocities for all eight double peaked infall sources (Table \ref{tab:infall_rates}).  The mass infall rate can then be determined using: 

\begin{equation}
\dot{M} = \frac{dM}{dt} \approx \frac{M}{t} = \frac{\rho V v_{\rm in}}{r_{\rm gm}} =\frac{4}{3}\pi n_{\rm H_2} \mu m_{\rm H} r_{\rm gm}^2 v_{\rm in}
\label{eqn:infall}
\end{equation}

\noindent where $\mu$ is the mean molecular weight ($\mu=2.35$), the geometric mean radius ($r_{\rm gm}$) is the unresolved circular radius of the HCO$^+$ emitting region derived from the beam radius and the beam filling factor ($r_{\rm gm}=\sqrt{f}r_{\rm beam}$), and $n_{\rm H_2}$ is the ambient source density. For seven of the sources in Table \ref{tab:infall_rates}, the ambient density was determined by either P97, Hofner et al. (2000), or Wang et al. (1993). We could not find the ambient density for the eighth source (G192.58). From this analysis, we determine mass infall rates ranging from 1$\times10^{-2}$ to 2$\times10^{-5}$ M$_{\odot}$ yr$^{-1}$. These values are slightly higher than those generally observed for low mass star forming regions, but are consistent with the accretion rates derived for high mass star forming regions by McKee \& Tan (2003).  Since outflow rates are orders of magnitude higher in high mass star forming regions (i.e. Beuther et al. 2002) it is not unreasonable to suggest that infall rates are also much higher in these regions.

The mass outflow rate for only one of these sources (Cep A) can be determined from the Wu et al. (2004) survey of high velocity outflows by dividing the mass in the outflow  by the kinematic age of the outflow.  We find the ratio of the mass outflow rate to the mass infall rate to be $\dot{M}_{\rm out}/\dot{M}_{\rm in}\approx16$. This value is only slightly higher than values seen in other high mass star forming regions (i.e. Behrend \& Maeder 2001). Also, models suggest that a mass equivalent to 20-30\% of the mass accreted onto a protostar is ejected as a wind (i.e. Pelletier \& Pudritz 1992, Shu et al. 1994), and that this wind entrains 5-20 times its mass in  the outflow (Matzner \& McKee 1999).

\subsection{Source properties derived from mapping G45.07}
\label{sec:g4507}

At a distance of 9.7 kpc, G45.07 is one of our furthest sources.  This source was known from previous observations to have multiple continuum sources (De Buizer et al. 2003, 2005). The three continuum sources were observed in the mid-Infrared (MIR) and all three fall within 6$''$ of our map center. A fairly young outflow has also been mapped at high resolution ($<3''$ synthesized beam) in CO and CS  towards this region (Hunter et al. 1997).  They observed a bipolar outflow with a position angle of -30$^{\circ}$ (east of north), as well as a red shifted absorption feature in their CS observations which they take to be indicative of infall.

\begin{figure}
\vspace*{6cm}
\includegraphics{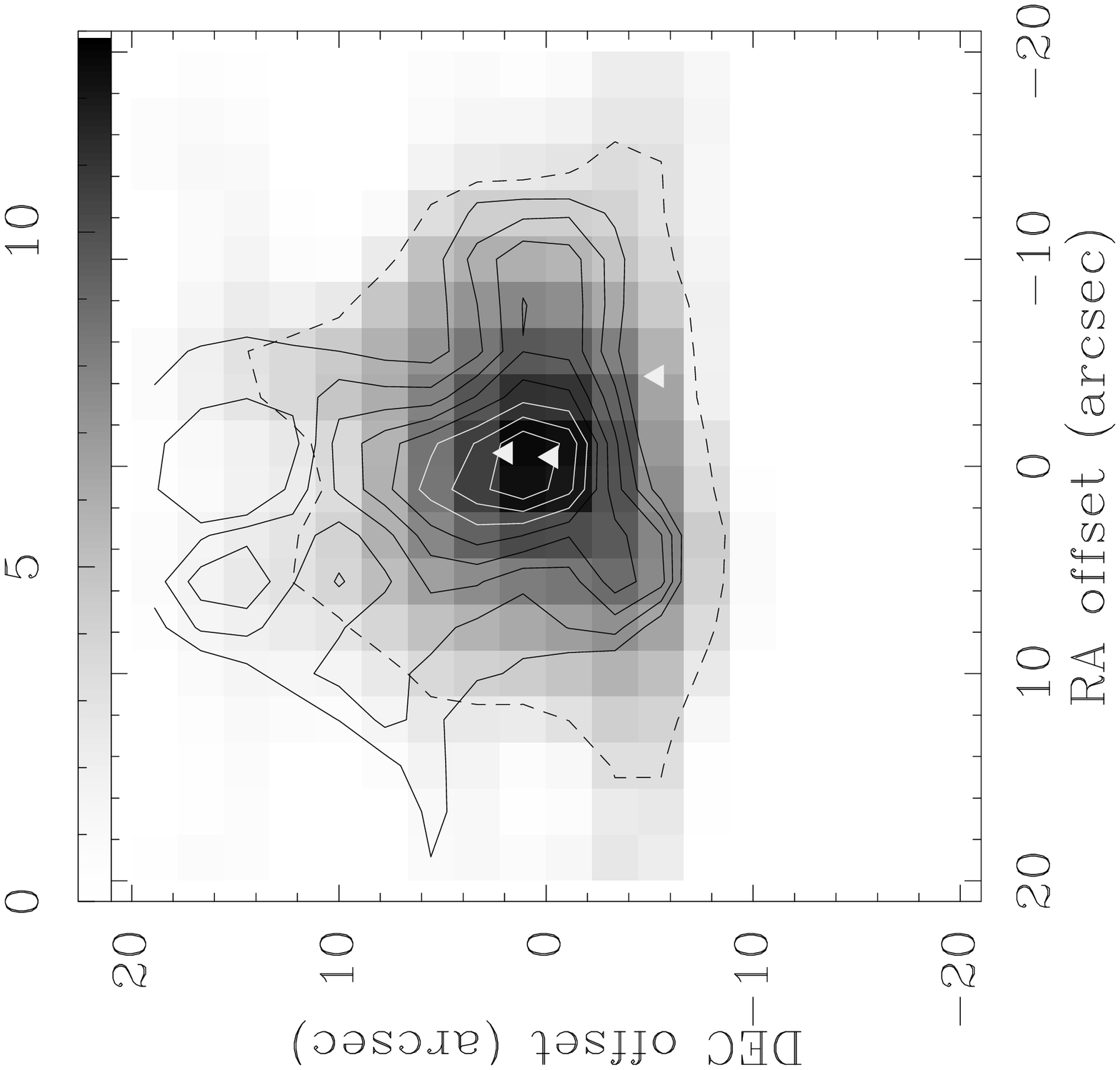}
\includegraphics{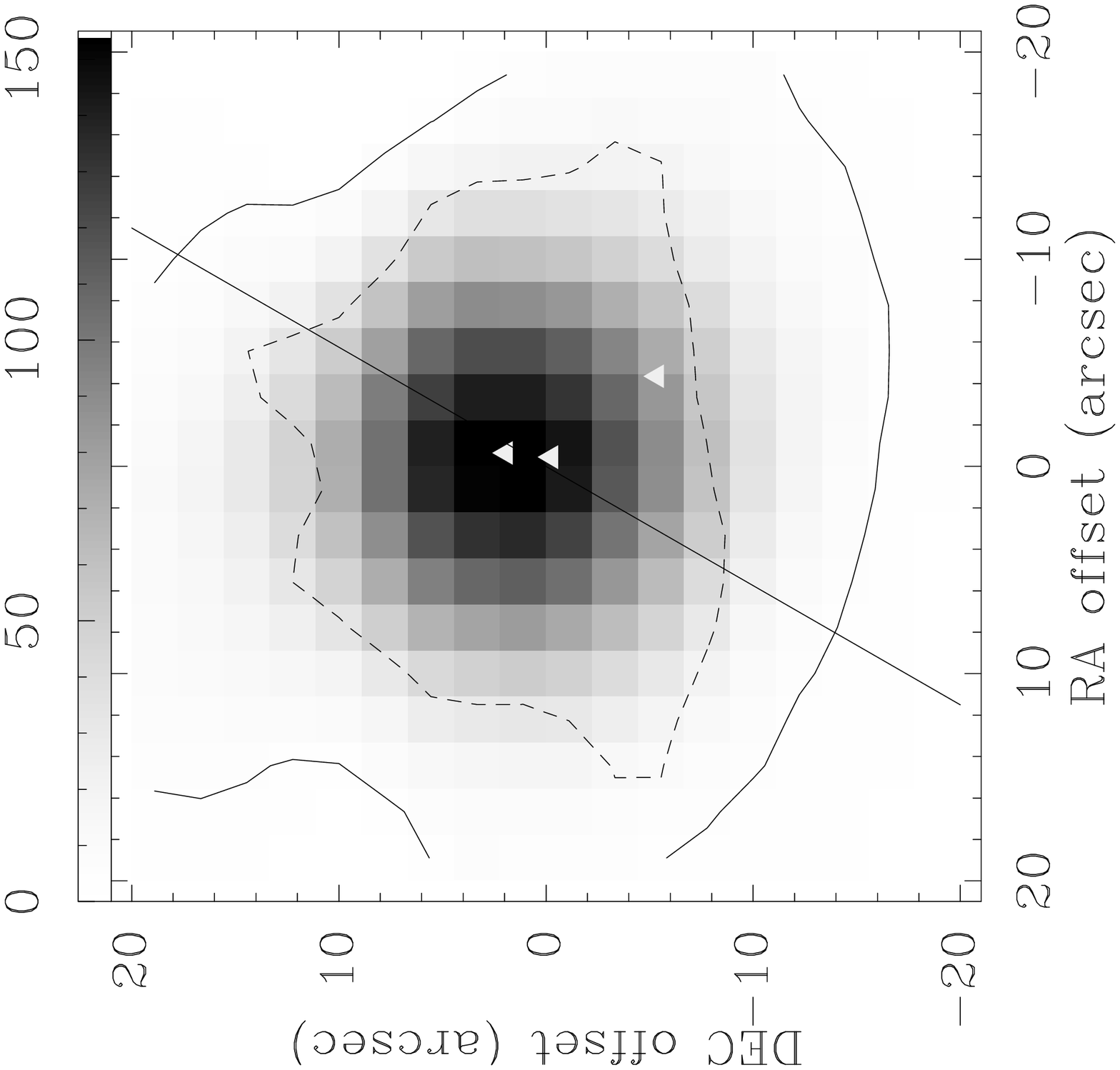}
\caption{{\bf Left:} SiO and H$^{13}$CO$^+$ towards G45.07.  The halftone scale represents the integrated H$^{13}$CO$^+$ emission, while the contours show the integrated intensity of SiO. The velocity range used to determine the integrated intensities is the same as the single pointing velocity range ($dv$), however the rms noise limits are those listed at the end of Section 2. The first SiO contour is 5$\sigma$ (3.2 K km s$^{-1}$), incrementing in steps of 2$\sigma$, with the same scale continuing into the white contours near the center.  {\bf Right:} HCO$^+$ integrated intensity towards G45.07.  The solid contour represents the 5$\sigma$ emission for the HCO$^+$ emission (3.5 K km s$^{-1}$), and the solid line represents the cut used for the PV diagram along the outflow axis (PA = -30$^{\circ}$) presented in Figure \ref{fig:PV}. For both panels, the dashed line represents the 5$\sigma$ H$^{13}$CO$^+$ emission contour (3.0 K km s$^{-1}$).  The three triangles represent the Mid IR sources detected by De Buizer et al. (2005), with the two points in the center corresponding to 230 GHz continuum sources detected at the Submillimeter Array (Klaassen et al. in prep).}
\label{fig:G45_maps}
\end{figure}

Due to the large distance to this source, we should be able to detect all of the emission associated with this source in a fairly small map.  In the left panel of Figure \ref{fig:G45_maps} we present a map of the SiO (contours) and H$^{13}$CO$^+$ (halftone) emission in this region. The right panel of  Figure \ref{fig:G45_maps} shows the HCO$^+$ emission. In both figures, the 5$\sigma$ (2.1 K km s$^{-1}$) H$^{13}$CO$^+$ emission contour is plotted as a dashed line to help guide the eye. The first contour for SiO in the left panel and only contour of HCO$^+$ in the right panel are also 5$\sigma$ (2.2 and 3.5 K km s$^{-1}$ respectively). The differences in the 5$\sigma$ contour levels for each tracer come from the different single channel rms levels between the two tunings, and the width of each line as given in Table \ref{tab:fits}.   Also plotted in both figures are the three MIR continuum sources observed by De Buizer et al (2003, 2005).

If we did not have the added information provided by this map, there would be two main conclusions we could draw from our single pointing observations towards this source. The first is that the enhanced blue emission in all three tracers suggests that our pointing is observing more of the blue shifted outflow lobe than the red.  Second, since we only have one peak in the spectrum of each tracer, there is only one source and we cannot classify it as infalling.

The left panel of Figure \ref{fig:G45_maps} shows contours of SiO emission superimposed on the H$^{13}$CO$^+$ halftone. With beam spacings of 5$''$, these maps are oversampled; however we note that much of the structure in the SiO emission is on scales comparable to the size of the JCMT beam.  For instance, the structure at $\Delta\alpha=-10''$, $\Delta\delta=0''$ is offset from the map center by more than the radius of our beam and could be independent from the emission at the map center.  There is also SiO emission at $\Delta\alpha=5''$, $\Delta\delta=15''$, which is more than a full beam away from the map center, and suggests that the SiO emission is more extended that the primary beam of our observations.  In fact, it appears as though there is a second SiO emission peak towards the upper left of the left panel of Figure \ref{fig:G45_maps}. Interestingly, there does not appear to be as much H$^{13}$CO$^+$ emission at this northern position. This comparison shows that the SiO and H$^{13}$CO$^+$ lines are tracing different gas populations in this region.  The excess SiO emission is offset from the map center in the same direction as the CO emission shown in Hunter et al. (1997) at much higher resolution.

The right panel of Figure \ref{fig:G45_maps} shows the HCO$^+$ emission for this region.  It appears that the HCO$^+$ emission extends much further than the SiO emission, suggesting it is tracing the larger scale envelope material. The line through the middle of this plot indicates the cut taken for the position-velocity (PV) diagram along the outflow axis as described by Hunter et al. (1997)

\begin{figure}
\includegraphics[scale=0.7,angle=-90]{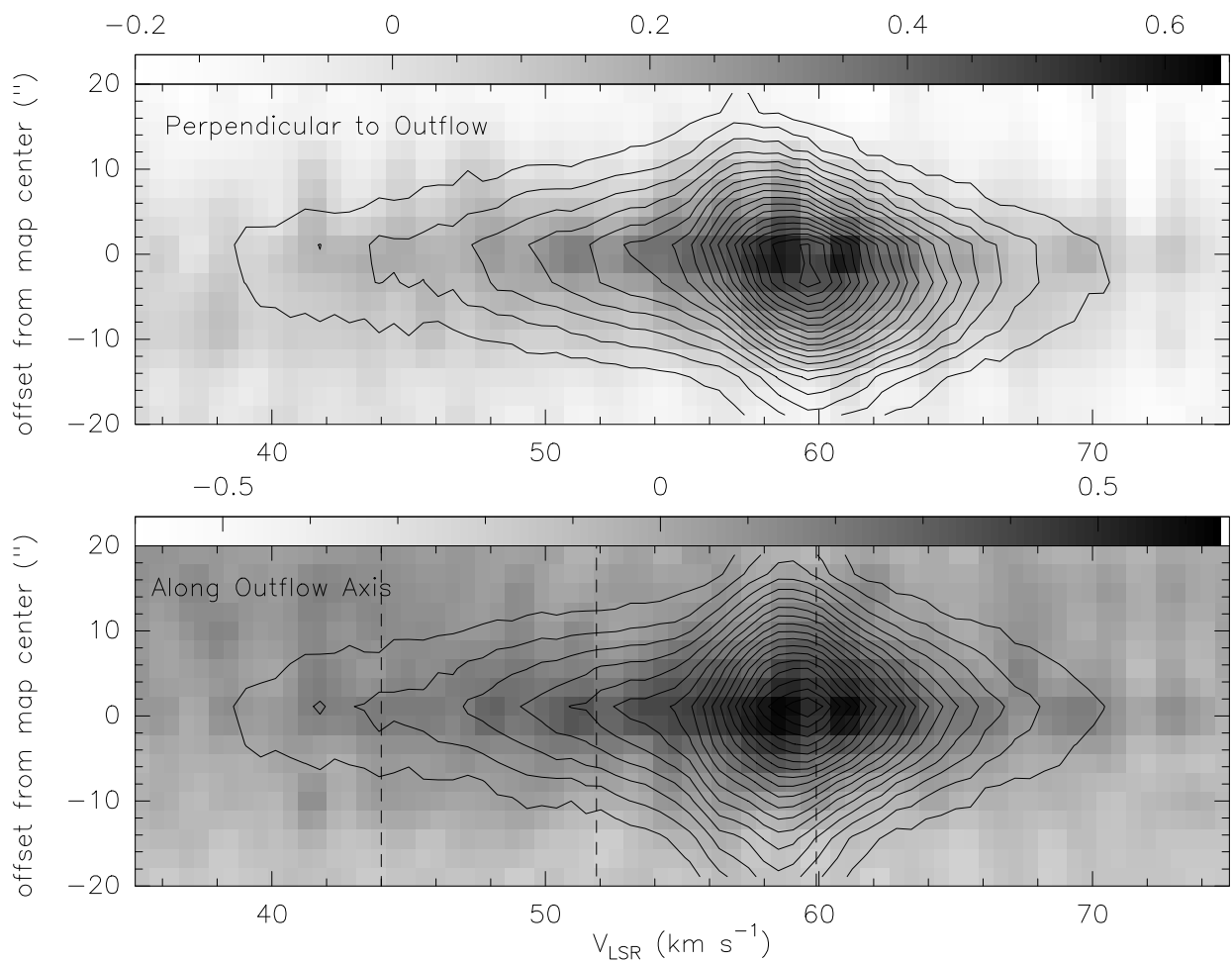}
\caption{Position-Velocity (PV) diagrams for SiO (halftone scale) and HCO$^+$ emission (contours) in G45.07 both perpendicular to (top panel) and along the outflow axis (bottom panel) as defined in Hunter et al. (1997).  These PV diagrams were taken at positions angles of 60$^{\circ}$ and -30$^{\circ}$ east of north respectively, with the cut for the bottom panel of this figure shown in the right panel of Figure \ref{fig:G45_maps}. The first contour for HCO$^+$ is 5$\sigma$, or 0.7 K since the rms noise limit for this map is 0.14 K, and the contours increase in increments of 5$\sigma$. The three dashed lines in the bottom panel show the peak velocities of the Gaussian fits to the HCO$^+$ spectra.}
\label{fig:PV}
\end{figure}

 The two panels of Figure \ref{fig:PV} show the PV diagrams for SiO and HCO$^+$ in our maps both perpendicular and parallel to the outflow axis defined by Hunter et al. (1997). Our single pointing HCO$^+$ spectrum  (Figure \ref{fig:Hdet_sio})  suggests we are observing more blue shifted outflow emission than red shifted emission; however, from our PV diagrams, we see that there is excess blue emission at all positions in our map. This excess blue emission cannot be due to outflow alone; instead, it could be due to an inherent velocity shift between the three continuum sources in our beam. We can, in fact, fit three Gaussian components to most of our HCO$^+$ spectra.  These Gaussians peak at velocities of 60, 52 and 44 km s$^{-1}$, with the peak temperature for each component decreasing with velocity.  The third component (at 44 km s$^{-1}$) could not be fit at all positions because it was intrinsically weaker than the other two peaks, and was lost in the noise towards the edges of the map.  It appears as though this third component might  be contamination from SO$_2$, which should occur at an apparent velocity of 41 km s$^{-1}$ (or -17 km s$^{-1}$ in Figure \ref{fig:Hdet_sio}).

Perpendicular to the outflow axis, the mean velocity of the HCO$^+$ line peak appears to shift from $\sim$ 58 km s$^{-1}$  at an offset of $+15''$ from the source center to $\sim 61$ km s$^{-1}$ at an offset of $-15''$ from the source (Figure \ref{fig:PV}). Given our velocity resolution (1.08 km s$^{-1}$) and spatial resolution ($15''$), is unclear whether this velocity shift is real.  If it is, it could indicate large scale ($\sim$ 1.4 pc) rotation within the core, on a much larger scale than would be expected for a rotating accretion disk.

\subsection{Correlations between Datasets}
\label{sec:discussion}

\begin{figure}
\includegraphics[scale=0.7,angle=-90]{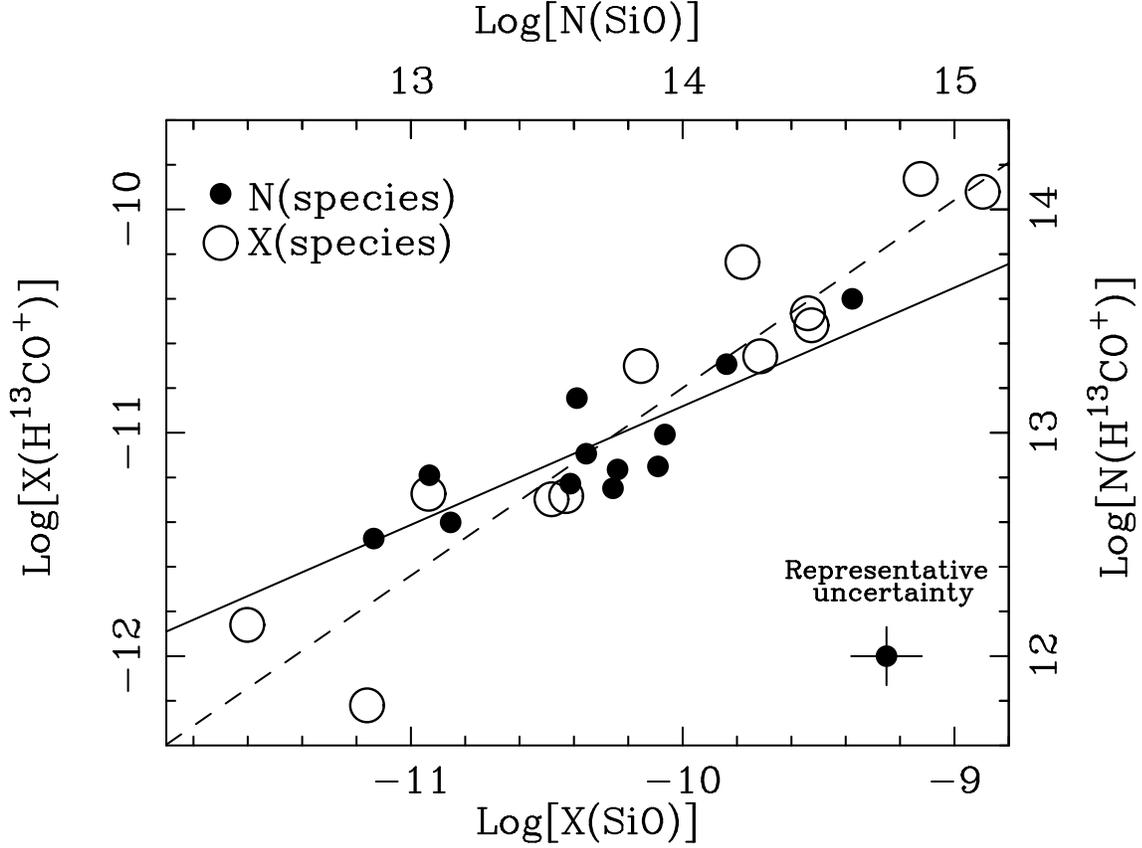}
\caption{The abundance of H$^{13}$CO$^+$ with respect to the abundance of SiO (open circles, dashed line of best fit) appears to increase faster than the respective column densities of these two species (filled circles, solid line of best fit). The equations of the two lines of best fit are: Log[X(H$^{13}$CO$^+$)]=(0.84$\pm$0.09)Log[X(SiO)]-(2.4$\pm$0.9), and Log[N(H$^{13}$CO$^+$)]=(0.5$\pm$0.1)Log[N(SiO)]+(5$\pm$1). This relationship could be due to HCO$^+$ being enhanced (similarly to SiO), as discussed in the text.}
\label{fig:abund}
\end{figure}

Previously, we discussed the reasons why we do not detect SiO towards a number of our sources, and have calculated the mass infall rates for the sources with double peaks in their HCO$^+$ emission, but we have not  yet discussed the correlations between the two species.  In Figure \ref{fig:abund} we plot the logarithm of the abundance of H$^{13}$CO$^+$ against the same quantity for SiO (open circles), as well as the column densities of both species (filled circles). The abundance of H$^{13}$CO$^+$ was calculated in the same manner as the abundance of SiO described above (using the CS column density from P97 and the CS abundance from S03).  The probability of obtaining these correlations if the data are, in fact, uncorrelated is 6$\times10^{-3}$ for the abundances, and 2$\times10^{-4}$ for the column densities.

As stated earlier, SiO is a well known shock tracer, and as such, an increased abundance of SiO would suggest more shocked material within our 15$''$ beam.  The (generally) infall tracing HCO$^+$ has been shown by some authors to be destroyed in strong shocks (i.e. Bergin et al. 1998, J\o rgensen et al. 2004). However,  Wolfire \& K\"onigl (1993) suggest that HCO$^+$ can be enhanced in regions with high energy shocks, where electron abundances are much greater. This enhancement in the electron abundance increases the formation rate of ions, and we suggest that this is responsible for  the H$^{13}$CO$^+$ abundance enhancement in our sources.  This  correlation between the abundances of H$^{13}$CO$^+$ and SiO suggests that HCO$^+$ and H$^{13}$CO$^+$ are not only tracing infalling gas, but also the outflowing gas as well.  This conclusion is supported by the strong, and broad, line wing emission detected in HCO$^+$ (See Section \ref{sec:hcop}).

HCO$^+$ over abundances have been seen in high mass star forming regions not included in this study like NGC 2071 (Girart et al. 1999) and Orion IRc2 (Vogel et al. 1984). In these two papers, the over abundances of H$^{13}$CO$^+$ are with respect to ambient cloud tracers such as CO and H$_2$, rather than the high density or shock tracers like the CS and SiO with which we are comparing our H$^{13}$CO$^+$  abundances. However, Viti \& Williams (1999a,b) show that HCO$^+$ is indeed over abundant with respect to CS in the gas surrounding HH objects, and Jim\'enez-Serra et al. (2006) also show that the abundance of H$^{13}$CO$^+$ can be enhanced with respect to SiO by up to a factor of ten in the same regions ahead of HH objects.

\section{Discussion and Conclusions}
\label{sec:conclusions}

Without maps of each region, it is impossible to tell how much of the HCO$^+$ emission in the line center and in the line wings is due directly to infall and outflow motions; however, based on the arguments we have presented above, we suggest that ten sources show infall motions, and all 23 source show outflow motions based on the HCO$^+$ line profiles. We have found evidence for recent outflow activity (SiO emission) in 14 out of our 23 sources.  Seven of these outflow and infall sources overlap. M17S, G192.6 and G240 appear to show only infall signatures and no SiO outflow signatures. They do, however,  appear to have HCO$^+$ outflow signatures of a minimum of 8$\sigma$. 

Detection of line wing emission in HCO$^+$ and the relationship between H$^{13}$CO$^+$ and SiO abundances described in the previous section suggest that while SiO is tracing outflow in most sources and HCO$^+$ is tracing infall in some sources, HCO$^+$ is also observable in the outflowing gas for all regions. 

 We find that the non-detection of SiO in nine of our sources is not due to beam dilution or larger average distances to the source, but possibly to older outflows for which the Si has likely either frozen back onto dust grains or evolved into SiO$_2$. For these sources, it appears that the accretion may have ceased, and the observed outflowing gas is a remnant of previous accretion.

We have found seven sources with SiO outflow signatures but no infall signatures in HCO$^+$.  This result could be due to a number of factors such as beam dilution of the infalling gas which masks the spectral line profile we would expect for large scale infall. 

It is possible that, as the outflow ages and widens, it may impinge on the region in which we could detect infalling gas.  For outflow cones oriented along the line of sight, the younger, narrower outflows would have infalling gas with large line of sight velocities and be likely to produce an observable infall signature. However, for the older outflows which have widened, the largest infall velocities will be in the plane of the sky, and unobservable at the resolution of the JCMT. However, this effect would not be as pronounced for outflows in the plane of the sky (such as G5.89, for which we do not see an infall signature).  It is difficult to assess the importance of this effect without detailed information on the outflow orientation in each source.

Thus, we suggest that some of these sources may have finished accreting, and what we observe are remnant outflows from a previous phase of accretion. This scenario was suggested by Klaassen et al. (2006) to explain the large scale outflow in G5.89, and this source is one of these seven sources with an SiO outflow and no apparent infall signature.

The seven sources which show recent outflow activity (those with SiO emission) and which appear to be undergoing infall  are suggestive of ongoing accretion beyond the onset of the HII region. If accretion is ongoing in the presence of an HII region, then it seems likely that this accretion flow may be ionized.  This ionized accretion scenario could be similar to low angular momentum accretion with high ionization as suggested by Keto (2006). Thus, from this survey, we suggest that both models of ionized accretion and halted accretion may be important in describing the evolution of a massive protostar (or protostars) beyond the formation of an HII region.

\acknowledgements

We would like to acknowledge the support of the National Science and Engineering Research Council of Canada (NSERC). We thanks the referee for helpful comments which improved the paper. P.D.K would also like to thank E. Keto and D. Johnstone for helpful discussions during the preparation of this manuscript.

\clearpage

\begin{deluxetable}{lllcccccc}
\tablecolumns{8}
\tablewidth{0pc}
\tablecaption{Observed Sample of Massive Star Forming Regions.}
\tablehead{
\colhead{Name}   &\multicolumn{2}{c}{Position (J2000)} & \multicolumn{2}{c}{RMS noise limit (K)}
&\colhead{V$_{\rm LSR}$}&\multicolumn{2}{c}{Distance}&\colhead{P92 Name\tablenotemark{a}}\\
\cline{2-3} \cline{4-5} \cline{7-8}\\
\colhead{}&  \colhead{RA} & \colhead{DEC}&  \colhead{347 GHz}& \colhead{356 GHz}	&\colhead{(km s$^{-1}$)}&\colhead{(kpc)} & \colhead{ref}} 
\startdata
G5.89		&18 00 30.3	&-24 03 58	&0.060	&0.111&	9	&2&1 & W28A2 (1)\\
G5.97		&18 03 40.4	&-24 22 44	&0.044	&0.069&	10	&2.7&5& \nodata\\
G8.67		&18 06 19.0	&-21 37 32	&0.068	&0.074&36	&8.5&2& 8.67-0.36\\
G10.47		&18 08 38.4	&-19 51 52	&\nodata\tablenotemark{b}	&0.100&67	&12&2& W31 (1)\\
G12.21		&18 12 39.7	&-18 24 21	&0.042	&0.076&24	&16.3&2& 12.21-0.1\\
M17S		&18 20 24.8	&-16 11 35	&0.044	&0.077&20	&2.3&5 & M17 (2)\\
G19.61		&18 27 38.1	&-11 56 40	&0.048	&0.065&43	&4.5&1 & 19.61-0.23\\
G20.08		&18 28 10.4	&-11 28 49	&0.044	&0.068&	42	&4.1&1 & 20.08-0.13\\
G29.96		&18 46 03.9	&-02 39 22	&0.044	&0.079&	98	&9&1 & W43S\\
G31.41		&18 47 33.0	&-01 12 36	&0.044	&0.061&	97	&8.5&1 & 31.41+0.31\\
G34.26		&18 53 18.5	&01 14 58	&0.047	&0.131&	58	&3.7&1& W44\\
G45.07\tablenotemark{c}		&19 13 22.1	&10 50 53	&0.044	&0.073&	59	&9.7&1 & 45.07+0.13\\
G45.47		&19 14 25.6	&11 09 26	&0.037	&0.063&	58	&8.3&6 & \nodata\\
G61.48		&19 46 49.2	&25 12 48 	&0.037	&0.048&	12	&2&1 & S88 B\\
K3-50A		&20 01 45.6	&33 32 42	&0.035	&0.077&	-24	&8.6&3 & K3-50\\
G75.78		&20 21 44.1	&37 26 40	&0.038	&0.068&	0	&5.6&1 & ON 2N\\
Cep A		&22 56 17.9	&62 01 49 	&0.031	&0.076&-10	&0.7&1 & CEP A\\
W3(OH)		&02 27 03.8	&61 52 25 	&0.026	&0.079&	-48	&2.4&2 & W3 (OH)\\
G138.3		&03 01 29.2	&60 29 12	&0.066	&0.073&-38	&3.8&1 & S201\\
G139.9		&03 07 23.9	&58 30 53	&0.058	&0.071&	-39	&4.2&1 & \nodata\\
G192.58		&06 12 53.6	&17 59 27 	&0.037	&0.053&9	&2.5&3 & S255/7\\
G192.6		&06 12 53.6	&18 00 26	&0.074	&0.098&9	&2.5&3 & S255/7\\
G240.3		&07 44 51.9	&-24 07 40	&0.035	&0.048&68	&6.4&4 & \nodata\\
\enddata
\tablenotetext{a}{Name given to source in Plume et al. (1992). When included in P97 and S03, these were the source names used.}
\tablenotetext{b}{Too much chemistry in spectrum to determine a reliable rms noise level.}\\
\tablenotetext{c}{The rms noise limits for the two maps of this source are 0.1 K and 0.14 K for the 347 GHz and 356 GHz pointings respectively.}
\tablecomments{The source velocity here is the reference velocity of each source, which has been removed from each spectrum. RMS noise levels are given in T$_{\rm MB}$. References:(1) Hanson et al. 2002, (2) S03, (3) Kurtz et al. 1994, (4) Kumar et al. 2003., (5) Churchwell et al. 1990, and (6) Hofner et al. 2000}
\label{tab:sources}
\end{deluxetable}

\clearpage

\begin{deluxetable}{lcc@{$\pm$}cccc@{$\pm$}cccc@{$\pm$}cc}
\tablecolumns{11}
\tablewidth{0pc}
\tablecaption{Integrated Intensities for SiO, HCO$^+$ and H$^{13}$CO$^+$.}
\tablehead{
\colhead{Name} & \multicolumn{4}{c}{SiO Properties}& \multicolumn{4}{c}{HCO$^+$ Properties} & \multicolumn{4}{c}{H$^{13}$CO$^+$ Properties}\\
\cline{2-5} \cline{6-9} \cline{10-13}
\colhead{}	&\colhead{T$_{\rm MB}$}&\multicolumn{2}{c}{$\int$T$_{\rm MB}dv$}& \colhead{$dv$}
	&\colhead{T$_{\rm MB}$}&\multicolumn{2}{c}{$\int$T$_{\rm MB}dv$} &\colhead{$dv$}
	&\colhead{T$_{\rm MB}$}&\multicolumn{2}{c}{$\int$T$_{\rm MB}dv$} &\colhead{$dv$}
} 
\startdata
G5.89 	&2.3		&75.3	&0.7	&120	&39.2	&690.0	&0.8 &	105	&8.9	&65.2 	&0.4 &35\\
G5.97 	&$<$0.1	&$<$0.3&0.1	&8 	&12.4 	&50.5	&0.2&	15	&0.6	&1.6 	&0.1 &8\\
G8.67 	&0.2		&1.6	&0.3	&13	&13.8 	&80.0	&0.3&	38	&5.6	&20.9 	&0.2 &10\\
G10.47	&1.0		&9.9&\nodata	&22	&7.7 	&79.3	&0.4&	35	&1.2	&9.2&\nodata &15\\
G12.21	&0.2		&2.5	&0.2	&25	&10.0	&93.8 	&0.3&	30	&0.8	&6.5	&0.2 &15\\
M17S  	&$<$0.1	&$<$0.7&0.2	&15	&16.2 	&79.0 	&0.2&	15	&2.9	&9.4	&0.2 &11\\
G19.61	&0.8		&15.9	&0.3	&45	&10.8	&162.2	&0.4&	55	&1.8	&16.1 	&0.2 &20\\
G20.08	&0.4		&6.9	&0.3 	&40	&6.5 	&75.0	&0.3&	40	&1.1	&9.7	&0.2 &25\\
G29.96	&0.5		&7.9	&0.3	&40	&23.6 	&160.4 	&0.3&	30	&3.8	&13.2	&0.1 &10\\
G31.41	&$<$0.1	&$<$0.7&0.2	&15	&1.6 	&6.8	&0.2 &	13	&$<$0.1	&$<$0.3 	&0.1 &3	\\
G34.26	&1.3		&26.0 	&0.4	&55	&27.8 	&228.2 	&0.6&	35	&4.3	&33.2 	&0.2 &25\\
G45.07	&0.7		&10.3	&0.3	&37	&15.0 	&169.4	&0.4&	45	&1.5	&11.2 	&0.3 &33\\
G45.47	&$<$0.1	&$<$0.3	&0.1	&12	&10.1	&59.4 	&0.2&	20	&1.3	&6.1 	&0.1 &12\\
G61.48	&$<$0.1	&$<$0.3	&0.1	&8	&10.8	&58.5	&0.1&	18	&1.5	&4.7 	&0.1 &8	\\
K3-50A	&0.3		&2.1 	&0.1	&16	&18.6 	&154.8 	&0.3&	25	&1.8	&10.6 	&0.2 &18\\
G75.78	&0.3		&2.2	&0.1	&13	&12.5 	&136.8 	&0.3&	45	&2.5	&10.3 	&0.2 &17\\
Cep A	&0.5		&7.3 	&0.2	&45	&23.2 	&234.7 	&0.4&	60	&4.5	&23.4 	&0.2 &28\\
W3(OH)	&1.4		&14.5 	&0.2	&50	&18.5 	&148.2 	&0.3&	30	&2.4	&11.6 	&0.1 &15\\
G138.3	&$<$0.2	&$<$0.3	&0.1	&2 	&5.0 	&10.6 	&0.2&	8	&0.3 	&0.5 	&0.1 &4	\\
G139.9	&$<$0.2	&$<$0.3&0.1	&11	&8.6 	&18.8 	&0.1&	6	&$<$0.2 	&$<$0.3 	&0.1 &2	\\
G192.58	&0.3		&1.3	&0.1	&8 	&14.5 	&84.2 	&0.2&	25	&1.1	&5.5	&0.1 &15\\
G192.6	&$<$0.2	&$<$0.6&0.2	&30	&12.0 	&54.1 	&0.3&	15	&0.8	&2.2	&0.2 &8	\\
G240.3 	&$<$0.1	&$<$0.4&0.2	&10	&7.3 	&59.1 	&0.2&	30	&0.4	&2.5	&0.1 &14\\
\enddata
\tablecomments{ For the sources in which we did not detect SiO or H$^{13}$CO$^+$, 3$\sigma$ upper limits on the brightness temperature and integrated intensities are given. }
\label{tab:fits}
\end{deluxetable}

\clearpage

\begin{deluxetable}{lccrccccc}
\tablecolumns{9}
\tablewidth{0pc}
\tablecaption{Observed and Derived Source Parameters.}
\tablehead{
\colhead{Name}  & \multicolumn{2}{c}{Detection}& $f$\tablenotemark{a}& \multicolumn{2}{c}{Column density ($\times 10^{12}$)}&\colhead{[SiO]/[H$_2$]\tablenotemark{b}} & \colhead{L$_{FIR}$} & \colhead{t$_{\rm outflow}$\tablenotemark{c}}\\
\cline{2-3} \cline{5-6}
\colhead{}	& \colhead{SO$_2$} & \colhead{Infall}	 & \colhead{} & \colhead{H$^{13}$CO$^+$} & \colhead{SiO} & \colhead{} & \colhead{Log(L$_\odot$)}& \colhead{(10$^4$ yr)}		
}
\startdata
G5.89		&Y	&N	&0.79	&39.7&	421.0   &-8.90		&5.25\tablenotemark{d}&0.2			\\
G5.97		&N	&N	&0.28	&1.0&	$<$1.7	&\nodata	&5.23\tablenotemark{d}&\nodata		\\
G8.67		&Y	&Y	&0.24	&12.7&	9.0	&\nodata	&5.70\tablenotemark{d}&\nodata		\\
G10.47		&Y	&Y	&0.16	&5.6&	55.4	&-9.53 		&6.26\tablenotemark{d}&\nodata		\\
G12.21		&Y	&N	&0.22	&4.0&	14.0	&-10.15		&6.17\tablenotemark{d}&\nodata		\\
M17S		&N	&Y	&0.33	&5.7&	$<$3.9	&$<$-10.73&	5.72\tablenotemark{d}	&\nodata		\\
G19.61		&Y	&Y	&0.22	&9.8&	89.0	&-9.72 		&5.42\tablenotemark{d}&\nodata		\\
G20.08		&Y	&N	&0.13	&5.9&	38.6	&-10.48		&4.86\tablenotemark{d}&\nodata		\\
G29.96		&Y	&N	&0.49	&8.0&	44.2	&-9.12 		&6.30\tablenotemark{d}&\nodata		\\
G31.41		&N	&N	&\nodata	&$<$0.2&$<$3.9	&$<$-11.89&5.45\tablenotemark{d}&\nodata		\\
G34.26		&Y	&Y	&0.58	&20.2&	145.0	&-10.43		&5.77\tablenotemark{d}&\nodata		\\
G45.07		&Y	&N	&0.32	&6.8&	57.6	&-9.54 		&6.15\tablenotemark{d}&4		\\
G45.47		&Y	&N	&0.21	&3.7&	$<$1.7	&\nodata	&6.04\tablenotemark{d}&\nodata		\\
G61.48		&N	&N	&0.23	&2.9&	$<$1.7	&$<$-10.80	&5.01\tablenotemark{d}&7		\\
K3-50A		&Y	&N	&0.40	&6.5&	11.7	&-11.60		&6.35\tablenotemark{e}&\nodata		\\
G75.78		&Y	&N	&0.25	&6.3&	12.3	&\nodata	&5.65\tablenotemark{d}&3.7		\\
Cep A		&Y	&Y	&0.47	&14.2&	40.8	&-9.78 		&4.40\tablenotemark{f}&0.2		\\
W3(OH)		&Y	&Y	&0.39	&7.1&	81.1	&-11.16		&5.12\tablenotemark{e}&\nodata		\\
G138.3		&N	&N	&0.11	&0.3&	$<$1.7	&\nodata	&4.57\tablenotemark{e}&17		\\
G139.9		&N	&N	&\nodata	&$<$0.2&	$<$1.7	&\nodata&4.82\tablenotemark{e}	&6	\\
G192.58		&Y	&Y	&0.32	&3.4&	7.3	&-10.94		&4.79\tablenotemark{e}&50		\\
G192.6  	&Y	&Y	&0.26	&1.3&	$<$3.4	&\nodata	&\nodata&\nodata		\\
G240.3		&Y	&Y	&0.16	&1.5&	$<$2.2	&\nodata	&\nodata&2.3		\\
\enddata
\tablenotetext{a}{HCO$^+$ beam filling factor.}
\tablenotetext{b}{SiO abundance relative to H$_2$ for sources with CS observations (P97) and abundance calculations (S03).}
\tablenotetext{c}{Kinematic age of outflow from Wu et al. (2004).}
\tablenotetext{d}{Far Infrared Luminosities modified from Wood \& Churchwell (1989).}
\tablenotetext{e}{Far Infrared Luminosities modified from Kurtz et al. (1994).}
\tablenotetext{f}{Far Infrared Luminosities taken from Evans et al. (1981).}
\label{tab:results}
\end{deluxetable}

\clearpage

\begin{deluxetable}{lc@{$\pm$}ccc@{$\pm$}c}
\tablecolumns{6}
\tablewidth{0pc}
\tablecaption{Infall Velocities and Mass Infall Rates.}
\tablehead{
\colhead{Name}		& \multicolumn{2}{c}{V$_{\rm in}$\tablenotemark{a}} &\colhead{$n_{{\rm H}_2}$\tablenotemark{b}} 	& \multicolumn{2}{c}{$\dot{\rm M}_{\rm in}$\tablenotemark{c}}\\
\colhead{}		&\multicolumn{2}{c}{(km s$^{-1}$)} & \colhead{(10$^5$ cm$^{-1}$)} & \multicolumn{2}{c}{(10$^{-4}$ M$_{\odot}$ yr$^{-1}$)}\\
}
\startdata
G8.67	&0.4	&0.1	& 1.8	&4&2\\
G10.47	&1.8	&0.3	& 7.2	&100&80\\
M17S		&1.4	&0.5	& 5.0 	&4 & 2\\
Cep A		&0.23&0.07	& 10.0	&0.17&0.07\\
W3(OH)	&0.06	&0.02	& 60.0	&3&1\\
G192.58	&0.8	&0.3	& \nodata 	& \multicolumn{2}{c}{\nodata}\\
G192.6	&0.9	&0.4	& 4.0	&2&1\\
G34.26	&1.5	&0.3	&3.6	&14&4\\
\enddata
\tablenotetext{a}{Infall velocities for sources with double peaked HCO$^+$ profiles.}
\tablenotetext{b}{Ambient densities taken from P97 except for: G34.26 (Hofner et al. 2000), M17S (Wang et al. 1993).}
\tablenotetext{c}{Mass infall rates.}
\label{tab:infall_rates}
\end{deluxetable}

\clearpage

\clearpage

\clearpage

\clearpage

\clearpage

\clearpage

\clearpage


\clearpage

\clearpage

\clearpage

\end{document}